\documentclass[creativecommons]{eptcs}
\providecommand{\event}{DSL 2011}

\title{Accurate Programming: \\ Thinking about programs in terms of
  properties\thanks{This work is funded by the Swedish KK Foundation,
    the Fulbright program, and the US NSF.}\ \thanks{This tutorial is
    contained in lecture notes entitled ``Accurate Programming'' by
    Veronica Gaspes, Rex Page, and Walid Taha, available under the
    Creative Commons 3.0 Unported License online at
    \url{accurate-programming.org}.}}

\author{Walid Taha \qquad Veronica Gaspes \institute{Halmstad University,
    Halmstad, Sweden}  \email{\{Walid.Taha,Veronica.Gaspes\}@hh.se}
  \and Rex Page \institute{University of Oklahoma, Norman, OK, USA}
  \email{page.ou.edu}}

\def\titlerunning{Accurate Programming}
\def\authorrunning{W. Taha, V. Gaspes \& R. Page}

\usepackage[latin1]{inputenc}
\usepackage{graphicx}

\usepackage{amsmath}
\usepackage{amssymb}
\usepackage{theorem}

\numberwithin{equation}{section}

\theoremstyle{plain}

\newtheorem{thm}{Theorem}[section]

\theoremstyle{definition}

\newtheorem{exercise}[thm]{Exercise}

\usepackage{color}
\usepackage{fancyvrb}

\usepackage{verbatim}

\begin{document}

\maketitle

% \begin{center}{May 2011
% }
% \end{center}

\begin{abstract}
Accurate programming is a practical approach to producing high quality
programs.  It combines ideas from test-automation, test-driven
development, agile programming, and other state of the art software
development methods.  In addition to building on approaches that have
proven effective in practice, it emphasizes concepts that help
programmers sharpen their understanding of both the problems they are
solving and the solutions they come up with.  This is achieved by
encouraging programmers to think about programs in terms of
properties.
\end{abstract}

\thispagestyle{empty}
\pagestyle{empty}

\newcommand{\indexed}[1]{\index{#1}#1}

\pagestyle{plain}
\section{Introduction}

Technical usage differentiates being accurate from being precise.
{``In the fields of science, engineering, industry, and statistics,
  the accuracy of a measurement system is the degree of closeness of
  measurements of a quantity to its actual (true) value. The precision
  of a measurement system, also called reproducibility or
  repeatability, is the degree to which repeated measurements under
  unchanged conditions show the same results.  Although the two words
  can be synonymous in colloquial use, they are deliberately
  contrasted in the context of the scientific method.''}
\cite{Accuracy}

Accurate programming is the idea that thinking about mathematical
properties of programs as we are developing them helps us produce
better programs.  We use the term {\em program} in the broadest sense
to include software, hardware, protocols, and algorithms in general.
With the growing success of what is known as property-based testing,
thinking about mathematical program properties has suddenly become
much easier and much more accessible than it ever was before.  These
notes introduce accurate programming using Scala as the programming
language for examples, and using the ScalaCheck library for specifying
and randomly checking program properties.

\subsection{Why are program properties such an important concept?}
In recent years the techniques for developing software and hardware
systems seems to have picked up pace dramatically.  This trend has
been seen in both broad-market methods (such as static typing, static
analysis, unit-testing, extreme programming, and more broadly,
test-driven development) as well as special-purpose, high-end methods
(such as model checking, other mathematical correctness methods, and
clean-room methods).  The upside of this progress is that we seem to
have both much better understanding of how to build such systems well.
The downside is that there has been an explosion in the number of
terms and concepts relating to the construction and analysis of
software and hardware systems.

The main point that these notes aim to get across is that the
mathematical notion of a {\em property} can play a central role in
helping us organize our understanding of many of these techniques as
well as how they relate to or differ from each other.  That is, the
notion helps clarify why certain recently developed techniques such as
property-based testing \cite{QuickCheck,Test} bear special promise for
improving the state of the art in building computational systems.

Thinking about program properties, especially with the aid of a
property-based testing tool, simultaneously
\begin{enumerate}
 \item reduces the number of defects in our code,
 \item enhances our understanding of our code,
 \item provides us with a powerful, practical way to gain real,
   hands-on experience with writing and understanding specifications
   of program properties, and
 \item actually makes it easier to go all the way to mathematically
   prove the correctness our programs, if and when an appropriate
   technology for doing that kind of thing is available.
\end{enumerate}
We all know that we program to change the world in some real way.  So,
why should we bother with trying to think about programs in a more
mathematical, more abstract way?  In addition to the fact that it will
allow us to write programs of significantly higher quality, it also has
the immediate benefit of simplifying a lot of concepts, and helping us
make a lot of connections between a wide range of concepts that may
have previously seemed disconnected.

As an example, let us try to see if thinking abstractly can give us
better intuitions about why programming is hard.  In abstract terms, ``to
change the world'' means to transform one point of its state space
into another.  As we program, we quickly build up complex
transformations that go far beyond simple intuitions.  To manage this
complexity, our most important tool is careful reasoning, or, put more
bluntly, our brain.  While programming and careful reasoning are
naturally very closely related, it is a curious fact that traditional
programming languages and models tend to obfuscate this relation.
Programming, we are often told, is about writing programs, and careful
reasoning is the realm of logic or mathematics.  Another example is an
idea that we hope these notes will drive home, which is that programs
not only have mathematical properties but they are also 
mathematical properties themselves.

Viewing programming from the perspective of properties provides a
unifying framework for understand strong connections between a range
of widely important computing concepts, including
\begin{itemize}
\item Use-cases and use-case analysis
\item Unit testing, test-driven development, and continual testing
\item Automatic test-case generation
\item Types, static type checking, and type inference
\item Mathematically stating and proving correctness
\item Contracts and dynamic monitoring
\end{itemize}

\subsection{Properties as a tool for the serious hacker}
Far too often, people talking about production programming seem to
think that hacking is a bad thing.  But hacking, in the sense of
experimental programming, is not only fun but also an extremely
useful activity.  To see that, however, we need to spell out what we
mean by the word and analyze the activity with some care.

Here, we use the term to mean the activity of sitting in front of the
computer, writing programs, running them, seeing what they do,
revising them, and repeating the process.  Surely, other people
will have other definitions, and we will not even argue that our
definition is superior or popular.  It's just what we mean by the
term here.

Hacking is a way of finding out if we can write a program to achieve a
certain functionality, and to accumulate knowledge about what the
language can do, what the machine or compiler running the language can
do, or what we as programmers may be able to code up.  When hacking is
done in the manner described above, the rapid cycle of program
modification and execution can provide us with a high-frequency stream
of new discoveries about the system that we are exploring.  This is a
big part of what makes hacking fun.  We are creating experiments,
running them, and very quickly learning a lot about a system that we
are interested in.

Of course, as an experimental process, hacking does have some weak
spots.  The good news is that the property-based view of programming
can actually help us fix these and turn hacking into an activity that
is not only enjoyable but also productive.  Here are some of the
things that happen during hacking that we do not tend to think about
very much.  The first is that older versions of programs tend to get
thrown away: we usually edit the same file in place.  This means that
we lose one way in which our thought and reasoning process could have
been documented.  But maybe the history of the code as it is being
edited is not the best record of our reasoning while hacking.  In
fact, often it is the tests that we run on these programs (as well as
the expected outputs that we usually never document) that are probably
the best record of our thinking process.  Thus, a more serious problem
is that we generally do not keep track of the tests that we run.  An
interesting side effect of this is that often newer versions of the
program that we are editing do not pass tests that older versions ran.
If we had somehow documented the old test cases, we would have
quickly realized that we made a mistake when we made some change as we
were editing the program.  This means that we will often have the
impression that we have tested the most recent version more extensively
than we actually have.  Worst of all, it is often the case that while
hacking we tend to use far too few test cases to really provide us
with any assurance about the functionality that we think our code
provides.

As we will see in this tutorial, thinking about program properties
will help us spell out explicitly the kinds of tests that we would
like to run on our programs, and using a property-based framework will
allows us to automatically test these properties using a large number
of test cases.  This means that we can hack much more effectively, end
up with well-documented and well-tested code, and be able to explain
the results of our hacking to others much more clearly and
convincingly.

\subsection{What you can expect to learn from this tutorial}
At a practical level, this tutorial aims to convince the reader that:
\begin{itemize}
 \item Testing really helps you debug your programs and get them right
 faster.  But, you may say, writing test cases can simply be tedious and boring.
 Using properties to capture how you expect your code to behave helps
 you write the code and helps you get a lot more test cases ``for
 free''.

 \item Properties are a key concept for expressing what a program is or
 should be doing.  They are a fundamental tool in any notion of
 {\em accurate} programming because until we have a clear idea of where
 we are going, it is hard to imagine how we can get there, not to
 mention check whether we are there or not.  Thinking about program
 properties makes testing much more intellectually satisfying.

 \item Testing is much easier and works much better when we do it starting
 from small pieces of code and work our way up to the full system.
 The ``bottom-up'' or``from-the-ground-up'' approach is the most
 effective approach to accurate programming.  It captures what can be
 viewed as a divide-and-conquer strategy to software quality.  It's
 hard to make high-quality systems from poor-quality ingredients.
 Interestingly, mathematically proving properties of smaller programs
 is also {\em much} easier.  So, if your goal is not only extensive
 testing but also to mathematically prove the correctness of your code,
 your best strategy is the same: write and check properties of pieces
 of your code from the ground up.

 \item Examples and use cases are crucial for effective software
 development and are your entry point into the world of properties.
 The process of developed examples and use cases starts both the
 process of developing your test suite and of capturing your thoughts
 about your program in terms of properties.  Just as with testing
 and mathematical analysis, you want to have these for every smallest
 part of your code.  Starting your development by writing
 representative examples and documenting them in an executable form in
 your code will save you a {\em lot} of headaches when you are
 checking or proving more sophisticated properties.

\end{itemize}
At the end of this tutorial, you can expect to be able to:
\begin{itemize}
\item Understand why programming is actually quite hard.  This is not
 an immediately obvious fact.  Worse, the man on the street is
 inclined to think that programming is not hard, and that it is just something
 that teenagers do in their spare time.  One of the most fundamental
 benefits from learning to think about programs in terms of their
 properties is that we begin to learn the vocabulary needed for
 reasoning about programs and their behaviors, and what makes some
 programs easier to reason about and others harder.

\item Interact effectively with ``customers'' and others interested in
 your programs.  Properties provide us with a way to think about our
 interaction with the ``customer'' for the code that we are
 developing, both in terms of soliciting specific use cases and in
 terms of developing a precise understanding of some of the customer's
 more nuanced expectations about the code that they want.

\item Know how to get your programs much closer to being amenable to
 mathematical proofs.  There are several reasons for this.  The first
 is that a program is much easier to prove correct when it does not
 have bugs.  Testing, especially property-based testing, can
 be very effective at finding bugs.  The other reason is that using
 property-based testing means you also develop the properties along
 with the code, thus ensuring that you have all the inputs that
 you need to start the process of mathematically proving that a
 program is correct.
\end{itemize}

\subsection{What you will NOT learn in this tutorial}
While the goal of this tutorial is to teach you to think about and
express mathematical properties of programs, it is focused on using
testing as a practical and lightweight technique for catching obvious
cases when these properties do not hold.  This process does catch
bugs, and does increase the chances that your programs are correct,
but it certainly is not a mathematical proof that the properties of your program hold.  For several reasons, this tutorial will not
explicitly address the issue of proving programs correct.  These
include the fact that there are already other texts elsewhere about
using tools such as the Z or VDM property languages to prove the
correctness of programs.  They also include the fact that traditionally
the cost of carrying out such proofs has been prohibitively high for
most businesses, in terms of both training their personnel to carry out
such proofs and the time and effort it takes to compete the
proofs themselves, as well as to keep them updated when the code
changes.  At the same time, when tools to automate the proof process
exist, they require even more training time.  With the property-based
testing approaches, we get essentially most of the benefits of formal
verification with dramatically less work.

\subsection{Practicalities:  Using Scala and ScalaCheck for exercise problems}

This tutorial is written with the intention that the reader will read
the notes and use Scala and ScalaCheck to perform the exercise
problems presented in the text.  Therefore, it is important that you
have Scala and ScalaCheck installed and running on your machine before
starting to go through the rest of the tutorial.

To use Scala and ScalaCheck \cite{ScalaCheck} you need to download the
compiler and some tools that follow with it. The best way is to go to
the official site for downloading
\begin{center}
\texttt{http://www.scala-lang.org/downloads}  
\end{center}
and follow the instructions for your system. You need to have Java 1.5
or later in your system. If you don't have it, there are instructions
on the scala download page.  All information and documentation about
Scala is under
\begin{center}
\texttt{http://www.scala-lang.org/}  
\end{center}
Together with Scala you get the Scala Bazaar System (\texttt{sbaz}). 
To start using it, follow the instructions in the first section of
\begin{center}
\texttt{http://www.scala-lang.org/node/93}  
\end{center}
Now, in order to install ScalaCheck, you just have to use \texttt{sbaz}
doing
\begin{verbatim}
sbaz update
sbaz install scalacheck
\end{verbatim}
All documentation and more detailed instructions are under
\begin{center}
\texttt{http://code.google.com/p/scalacheck/}  
\end{center}
In order to test your installation, you can write a Scala program
\begin{verbatim}
import org.scalacheck._
import org.scalacheck.Prop._

object Hello{
  val trivial = Prop.forAll((n:Int) => n==n)
  def main(args: Array[String]){
    println("Hello " + args(0))
    trivial.check
  }
}
\end{verbatim}
save it in a file \emph{HelloWorld.scala} and then compile and run
with
\begin{verbatim}
prompt> scalac HelloWorld.scala
prompt> scala Hello someone
Hello someone
+ OK, passed 100 tests.
prompt>
\end{verbatim}
If you have any problems running this program, it could be the case
that sbaz did not find the latest release of ScalaCheck. Then you have
to do some work by hand! Under the directory
\begin{verbatim}
.../scala-2.9.0.final/lib
\end{verbatim}
place the file you find under
\begin{center}
\texttt{http://scalacheck.googlecode.com/files/scalacheck\_2.9.0-1.9.jar}  
\end{center}
with the name \emph{scalacheck.jar}.

\section{A Simple Function}

   Consider the following programming task: write a function that
   takes two integers and returns the larger of the two.  Even though
   this is a simple problem, it can be used to illustrate a wide range
   of best practices and concepts relating to software development.

\subsection{Examples as a communication tool}

   Mastering accurate programming requires awareness of several facts
   that are quite simple but at the same time of tremendous importance
   in practice.  One of these facts is the exceptionally important
   role that {\em concrete examples} play in effective communication.
   A concrete example is one in which all the details have been spelled
   out, and nothing is left to imagination or intuition.
   Here, we also use the word {\em communication} in the broadest
   sense to include writing a letter, giving a speech or presentation,
   discussing an insurance policy, or talking about computer programs.

   There are several reasons why concrete examples are important.  The
   first is that as humans we experience the world primarily through a
   small set of basic senses.  Yet we think about the world and
   communicate about the world in much more general terms than what
   our primary experiences are made of.  Our beliefs about primary
   experiences are much easier to communicate than more abstract
   thoughts about them.  For example, it is much easier to reach an
   agreement about what is a ``hot tea pot'' than what is a ``pretty
   tea pot'', because the first relates directly to our senses, while
   the latter is a much more sophisticated and context-dependent notion.
   The same kinds of difficulties arise when we try to communicate
   about programs.

   The second is that using concrete examples reduces our reliance on
   imagination or intuition, both of which can actually be hard to
   communicate, especially about complex objects such as programs.
   Programs are generally deceptively more complex than they appear at
   first.  This makes it particularly important to spell out concrete
   examples that communicate concrete facts about how we expect such
   programs to behave.

   The third and final reason is that it is too often the case that
   programs are built with too few test cases.  Making a habit of
   starting the process of creating programs by first creating
   concrete examples is not only a good way to start understanding the
   problem that we are trying to solve but also to ensure that there
   will always be a minimal set of test cases that we can easily use
   in the future to make sure that our program continues to work
   correctly as we update it, upgrade it, modify it, or attempt to
   improve it in any other way.

\subsection{Use cases, use-case analysis, and test-driven development}

   The reader may be somewhat surprised by the emphasis we put on this
   idea.  However, it is the key idea behind important methods in
   software engineering, such as Use-case Analysis \cite{UseCase} and
   Specification by Example \cite{ByExample}.  The idea of using
   examples is very simple, but experience repeatedly demonstrates that
   they are one of the most important ingredients of successful
   program development.  Not only are examples useful for accurately
   understanding the functionality that we are being asked to program,
   but they are also extremely helpful in checking that we have attained
   and maintained the functionality that we think we have attained.
   This observation is the basis for widely successful methods of {\em
     test-driven development}, including techniques such as {\em unit
     testing} and {\em extreme programming}.  Test-driven development
   promotes the idea of starting development with building an
   appropriate set of tests for the functionality we want to develop.
   Unit testing pushes this idea a bit further to emphasize that it
   is most effective when applied at a fine granularity, so that tests are
   made when implementing the smallest possible unit of functionality.
   This idea of improving the quality of software from the ground up
   is very important, whether we simply want to make sure that we have
   a high-quality test suite for our code or we want to go all the
   way to prove our programs correct.

\subsection{Examples as properties}

   It is easy to put these ideas to work.  For the most part, it is
   just a matter of making it into a habit.  We will do this
   consistently in this tutorial.  For the \verb|max| problem, we may
   be able to solicit the following concrete use-cases:

\begin{Verbatim}
      max (1,5)  == 5
      max (1,1)  == 1
      max (3,2)  == 3
      max (3,-1) == 3
      max (1,-3) == 1
\end{Verbatim}

   With these concrete examples in hand, we get a more specific idea
   of what is required.  The last example above is actually
   particularly useful, because it excludes the possibility that the
   customer actually meant us to return the value with the greatest
   {\em magnitude} rather than simply the greatest value.  Naturally, language
   is very expressive, and much of its expressivity comes from the
   fact that the meaning of any one sentence can be highly context
   dependent.  Concrete examples can be very helpful in making sure
   that we are not making any false assumptions about the context.

   We can now confidently write out code for an implementation of this
   function:

\begin{Verbatim}
  def max (x : Int, y : Int) : Int = if (x>y) x else y
\end{Verbatim}

\subsection{Use cases as properties}

  ScalaCheck is a Scala library that allows us to explicitly specify
  and test a wide range of program properties, including specific test
  cases.  To capture the behavior expressed in the above examples, all
  we have to write is the following:

\begin{Verbatim}
  property ("Use cases for max") =
      (max (1,5)  == 5) &&
      (max (1,1)  == 1) &&
      (max (3,2)  == 3) &&
      (max (3,-1) == 3) &&
      (max (1,-3) == 1)
\end{Verbatim}

  This statement is executable test code.  When we run the above code
  (both definition and property statement), we get the following
  output when we run our program:

\begin{Verbatim}
  + SmallExamples.Use cases: OK, proved property.
\end{Verbatim}

  Not only is this a comforting confirmation that our code at least
  works on the examples that were discussed with our customer, it
  means we have an executable test that we can keep in our test
  harness so that we can always easily detect if any future changes
  to the code break functionality that we had already gotten right.
  The idea of keeping such test cases around and treating them as an
  integral part of the code that we developed is an essential part of
  test-driven approaches to software design.

  At this point we can stop, declare victory, and hand back our code
  to the customer and say that we are done.  While this could be a
  reasonable thing to do for a small function such as this one, for
  many problems we may want to study our own code a bit further to
  make sure that we really understand what it is doing.  Reading the
  code and giving it to other people to read is always useful.  But
  this short example also suggests that we and the other readers may
  feel that we don't necessarily know everything about the primitive
  operations used in this one-line program, so it is useful to
  produce more test cases to test some general properties that we
  expect to hold for this function.  An example of such a property
  could be that we expect that this function is symmetric-that is, it
  produces the same answer even if we switch the arguments around.  We
  can use concrete numbers to test this idea, but it would be even
  better if we can just write that down in a more general way.

\subsection{Universal quantification}

  Mathematics provides a great way to make but still very precise
  statements like the ones we are looking for.  For example, we can
  express the symmetry property that we require here as saying that we
  want that for all x and y that are Int values it will be the case
  that max(x,y) produces the same number as max (y,x).  Using more
  concise mathematical notation, we can write this statement as
  follows:
\[
  \forall x, y \in {\tt Int}.\;\;{\tt max}(x,y) = {\tt max}(y,x)
\]
  Technically, in a statement like the above we are leaving it
  implicit where the definition of {\tt Int} and {\tt max} are being
  looked up from, but the formulation above captures the gist of what
  we are trying to get at.

\subsection{Random testing of universally quantified properties}

  A universally quantified property such as the one above can be
  expressed as follows:

\begin{Verbatim}
  property("Symmetry") = 
   forAll ((x:Int, y:Int) => 
           max(x,y) == max(y,x))
\end{Verbatim}

The above statement is just a Scala program which makes some reference
to primitives defined in the ScalaCheck library.  When we run this
program, the system comes back and gives us a more interesting
response than the first one:

\begin{Verbatim}
  + SmallExamples.Symmetry: OK, passed 100 tests.
\end{Verbatim}

  Here, ScalaCheck generated 100 different pairs of values for x and y
  and used them to test the validity of our property.  ScalaCheck gets
  the hint that we will need it to generate some test integers for us
  when it sees the ``forAll'' operator.  The convenience of the forAll
  operator comes from the fact that it can alleviate the need for us to
  actually come up with 100 different pairs of numbers to write down
  such a test, not to mention writing them down explicitly.  Of course,
  there are many occasions when it is useful for us to write down
  individual test cases, but there are also many occasions when it is
  useful to have them be generated automatically.  In this case, it
  is also useful for us to document the symmetry property in our code,
  and in a manner that can be easily executed whenever we want to
  check the correctness of our code (which is virtually any time we
  make any change to it).

  We can continue to study our one-line program by considering other
  properties that it may have.  In this respect, it can be useful to
  think about how such a function could be used.  For example, we may
  use this two-argument max function to keep track of the largest
  number that we have seen as we go down a list.  From this point of
  view, if the next number we look at is the same as the maximum
  number we have seen so far, we'd like to keep the maximum the same.
  The following property tells us that the max function can do this
  for us automatically:

\begin{Verbatim}
  property("max(x,x)=x") = 
   forAll ((x:Int) => 
           max(x,x) == x)
\end{Verbatim}

  This last property illustrates how understanding the properties of
  the code that we have written well can often help us fully appreciate
  the behavior it provides and take full advantage of its
  functionality to simplify the way we use it.

  Another important property of this function is that the result it
  returns is an upper bound for the two values that it takes as
  argument.  This property can be expressed as follows:

\begin{Verbatim}
  property("Upper bound (2)") = 
   forAll ((x:Int, y:Int) => 
            x <= max(x,y) && 
            y <= max(x,y))
\end{Verbatim}

  Note that if we take into account the first property that we
  stated about max, symmetry, then the property above can actually
  be simplified to consider only one argument.  That is, when we
  have symmetry the property above is implied by the following,
  more concisely stated property:

\begin{Verbatim}
  property("Upper bound (1)") = 
   forAll ((x:Int, y:Int) => 
            x <= max(x,y))
\end{Verbatim}

\subsection{Properties that completely characterize a function}

  Often, it is useful to consider whether we have been able to express
  the simplest property that fully characterizes the behavior of the
  code that we wrote.  Interestingly, the above property (and, in
  fact, even all the properties that we have written above combined)
  do not fully characterize our max function.  In particular, the last
  property says that the result of max is {\em an} upper bound, and not
  necessarily a particular upper bound.  In fact, max computes the
  {\em least} upper bound of two numbers, in the sense that max produces
  the {\em least} number satisfying the above property.  This additional
  constraint can be expressed as follows:

\begin{Verbatim}
  property("Least upper bound") = 
    forAll ((x:Int, y:Int, u:Int) => 
            (x<=u && y<=u) ==> max(x,y) <= u)
\end{Verbatim}

  Together, these two seemingly very generic properties actually fully
  characterize the intended behavior of the maximum function.

\begin{exercise}  Consider the following function:

\begin{Verbatim}
  def mad_max (x : Int, y : Int) : Int = 
    if (x==42 && y==42) 43 
       else if (x>y) x else y
\end{Verbatim}

  Which of the properties considered above would break if we used this
  function in place of max?  First, write down your answers when you
  check the properties by hand.  Then, run all the tests using this
  function, and note any differences between what you expected and
  what you got from running the tests.
\end{exercise}

\section{Programs as a Special Type of Properties}

  It is instructive to note that certain properties can be equivalent,
  in the sense that they capture the same concept or behavior.  It is
  also useful to note that programs themselves can be viewed as
  properties.  To see this in the case of the max example, we can
  consider the following property:

\begin{Verbatim}
  property("Our implementation (A)") = 
    forAll ((x:Int, y:Int) => 
            if (x>y) max(x,y)==x else max(x,y)==y)
\end{Verbatim}

  It looks remarkably similar to the way we have implemented the max
  function.  The body of the if statement is not quite the same, but
  this is only a cosmetic difference.  We can express the same
  property as follows:

\begin{Verbatim}
  property("Our implementation (B)") = 
    forAll ((x:Int, y:Int) => 
            max(x,y) == (if (x>y) x else y))
\end{Verbatim}

  This property essentially gives us the code for our implementation.
  This is a very useful connection between properties and programs,
  and there are numerous situations in which converting one property into
  another equivalent property that has a different form can be a very
  useful method for deriving programs that implement the properties
  that we are interested in.  Note, however, that this generally
  requires that we have spelled out a set of properties that we want
  our program to have.  In many instances, however, we need to both
  develop the code and spell out its properties at the same time.  It
  is also interesting to note that this means that code/property code
  design is, in a sense, just property design.  The useful distinction
  in that case is that we are generally cross-testing two things
  against each other, and even though they are both properties, we are
  generally cross-testing different properties against each other.

\subsection{Properties as relations}

  It is reasonable to consider whether there are certain characteristics
  that make some properties closer to being programs than others.  One
  way to draw this distinction is to recognize that properties are
  generally mathematical relations between different sets.  For
  example, in the case of max, we wrote many properties that relate
  two or three sets of integers at the same time.  Intuitively, we can
  view properties as executable when they provide relations that are
  actually functions (which is a special kind of relation) from
  elements of a set that we consider to be the input to the
  computation into finite elements of another set that we
  consider to be the output of the computation.

\subsection{Programs as functions}

  Often an implementation is the best specification that we can
  express for a function.  This observation has a very real and
  concrete practical implication, which is that for some programs it
  is hard to write a specification without spelling out the code of
  the program itself.  This is a mixed blessing.  On the one hand, it
  means that we (either set out to or discover after the fact that
  what we just did is to) first write the code, and then we start
  thinking about other, ``extra-computational'' properties about it that
  we may want it to have.  This can seem strange at first, but it is
  something that can arise naturally in many situations, and that often
  produces very useful programs to have around.  All it means is that
  the program that you are describing has computational content that
  can be elegantly described, and you managed to find it.

  On the other hand, this deep observation should be approached 
  with care.  In particular, it does not mean that {\em all} programs are 
  elegant descriptions of their functionality.  Quite the contrary.
  Given that most programs in the world are buggy, it is highly unlikely
  that they are an elegant description of anything, or that there are
  any other properties that characterize their behavior and that could
  be viewed as elegant.  It only means that particular programs also 
  happen to be great specifications of their own behavior.  Examples
  of such programs include definitional interpreters for programming
  languages and insertion sort as a prototype for sorting.

\subsection{Functions as ``The Reference Implementations''}

  Usually, such self-evident programs are not simultaneously the best
  implementations in terms of performance.  Often, one still has to do
  a substantial amount of work to go from an implementation that is a
  nice specification to an implementation that is efficient.  For
  example, if we are talking about languages, compilers are generally
  more efficient implementations than interpreters.  If we are talking
  about sorting, there are numerous implementations of sorting that
  are much more efficient than the elegantly described insertion sort.
  Yet, in all of these cases, the self-evident or reference
  implementations are an invaluable tool for developing
  and testing the correctness of more efficient implementation.  For
  novel algorithms, however, the challenge lies in producing the first
  such reference implementation in the first place.  Thus, emphasis on
  code vs. property development while developing an algorithm that was
  not previously specified varies dramatically from the emphasis when
  we already have a reference implementation and we are trying to
  build a more efficient one.

  It is also useful to note that having the programming language and
  the property language be syntactically close is also a mixed
  blessing.  On one hand, it can facilitate turning an expression from
  being a program into a property and vice versa.  On the other hand,
  the intuitive meaning of the expression can also change in subtle
  ways when we do that, depending on the context.

\section{Functions on Numbers}

  The {\tt max} functions illustrate that even a one-line program can
  have interesting properties and deserves a reasonable degree of
  analysis.  Maybe more interesting than the fact that it is a one-  line program is the fact that it only involved arithmetic
  (comparisons) and a conditional statement.  In practice, many
  interesting programs involve iteration or recursion, and as a result
  both perform more computation per line of code and also exhibit more
  interest properties.

\subsection{An iterative program}

  As a simple example of a problem that requires iteration, consider
  the following problem: Write down a function that takes one
  argument, call it n, and computes the sum of the numbers counting up
  from 0 to~n.

  Before we start writing out examples, it is useful to
  note that there is implicit information in this problem
  statement that is useful to spell out.  Because the problem
  says ``count up from 0 to n'', it is reasonable to assume
  that our customer is assuming that the number n is either more
  than or (at least) not less than 0.  This also means that we 
  are left with the task of determining what the function should 
  do if the input is less than 0.  Now we can start creating
  some examples to convince ourselves that we understand the
  function required.  We will write it as a property so that it
  it is easy for us to just run the test when we have written
  the code.

\begin{Verbatim}
  property ("Use cases for sum") =
      (sum (-1) == 0) &&
      (sum (0)  == 0) &&
      (sum (1)  == 1) &&
      (sum (2)  == 3) &&
      (sum (3)  == 6) &&
      (sum (4)  == 10)
\end{Verbatim}

  If we are in a hurry, we can decide that we have satisfied the
  requirements of due diligence, having written out some examples, and
  then just write down an implementation of this simple function as
  follows:

\begin{Verbatim}
  def sum (n:Int) : Int = {
   var temp = 0
   for (i <- 1 to n)
     temp = temp + i
   temp
  }
\end{Verbatim}

\subsection{Extracting general properties from use cases}

  If we have a little more time, we can productively think a bit more
  carefully about this problem and about the code.  In fact, just by
  listing these examples, an interesting pattern emerges.  It's
  actually easy to compute the next value in this list of examples
  when we start from 0 (or below) because the result is 0 until we
  reach 1, and from that point on the result is the argument added to
  the result of the previous one.  For many problems in which we take a
  positive integer as input, this is an extremely helpful pattern to
  recognize when we want to write out code that solves a certain
  problem.  The two parts of what we have observed here can be
  expressed as the following two properties:

\begin{Verbatim}
  property ("Non-positive (A)") =
    forAll ((n:Int) =>
            (n<1) ==> (sum(n) == 0))

  property ("Positive (A)") =
    forAll ((n:Int) =>
            n>=1 ==> (sum(n) == n + sum (n-1)))
\end{Verbatim}

  These two properties are quite interesting because they fully
  characterize the sum function.  Not only that, but they can also be
  naturally rewritten in a step-by-step manner to {\em guide} us to
  executable code that captures the intuitions that we drew from the
  examples.  Granted, this is an extremely simple function that we are
  being asked to write, but it is useful to reflect on the process of
  writing, as it were, in ``slow motion''.  And the point is not that
  this is precisely how all problems should be approached, but rather
  that the connections between all of these different views of the
  problems exist, and that these different views are in fact
  themselves properties with strong connections among them.

\subsection{Reasoning about properties}

  The two properties above can be rewritten to look a bit more like
  code and less like a mathematical property.  In particular, we can
  rewrite them into :

\begin{Verbatim}
  property ("Non-positive (B)") =
    forAll ((n:Int) =>
            if (n<1) sum(n) == 0 else true)

  property ("Positive (B)") =
    forAll ((n:Int) =>
            if (n>=1) sum(n) == n + sum (n-1) else true)
\end{Verbatim}

  All we did here is replace any implication of the form
  \verb|A ==> B| into an \verb|(if (A) B else true)|.  Having true in the
  else branch of the if is consistent with the way mathematical
 implication works.  When the thing before the implication is not
 true, we can have anything after the implication and the whole
 property is still true.

  Next, we note that these conditions in the two if statements in each
  property are complementary.  In fact, we can rewrite the first
  property so that there is even more alignment between the two
  properties, and so that they can be combined into one:

\begin{Verbatim}
  property ("Non-positive (C)") =
    forAll ((n:Int) =>
            if (n<1) sum(n) == 0 else true)

  property ("Positive (C)") =
    forAll ((n:Int) =>
            if (n<1) true else sum(n) == n + sum (n-1))
\end{Verbatim}

  Only the second property was changed, and all that was done is to
  replace the condition \verb|(n >= 1)| with the dual condition
  \verb|(n < 1)| and then flip the else and then branches.  Note that
  it is easy to see that both properties have the form of an if
  statement with exactly the same condition.  Furthermore, each one
  has alternatively ``true'' in one branch and a more interesting
  condition in the other branch.  In fact, true is the least
  interesting condition to have in any property, because it is a
  trivial condition that makes no statements about any variables by
  itself.  Thus, what we can do is to combine both properties into
  one, keeping only the interesting branch of each property in the
  result.

\begin{Verbatim}
  property ("Sum (A)") =
    forAll ((n:Int) =>
            if (n<1) sum(n) == 0 else sum(n) == n + sum (n-1))
\end{Verbatim}

  At this point, we are very close to having the exact code for an
  executable program for computing the sum.  All we need is to note that
  both branches start with the expression sum(n) ==, and to
  reformulate our property one more time to take this equality
  {\em outside} the if statement:

\begin{Verbatim}
  property ("Sum (B)") =
    forAll ((n:Int) =>
            sum(n) == (if (n<1) 0 else n + sum (n-1)))
\end{Verbatim}

  Not only does this leave us with a shorter description of the same
  property, but we also have an equality test where on one side we have
  just the term sum(n), and on the other side we have an expression
  that can be viewed as a perfectly valid inductive definition of
  sum(n).  In particular, the if statement simply returns 0 for all
  non-positive values, and for positive values, it can compute a
  result by adding n to sum(n-1).

  Our original sum function, in fact, passes all the properties that
  we have seen so far.  These tests actually take a substantial amount
  of time to run, but they all pass.  We could indeed declare victory
  at this point and either go back to the customer and hand over our
  code or start thinking about other things, like optimizing our
  implementation.

\begin{exercise}

  The following definition for sum can be viewed as more efficient than
  the one above:

\begin{Verbatim}
  def sum2 (n:Int) : Int = if (n<1) 0 else (n+1) * n / 2
\end{Verbatim}

  Use ScalaCheck to check whether or not it satisfied all of the
  properties discussed above.
\end{exercise}

  But we can also continue to study our problem and its solution by
  considering more properties that we expect to hold about the
  function.

\subsection{Why working with numbers requires special care}

  Often, continuing to think carefully about the properties that
  should hold for a given program requires stepping back to identify
  what should be the most obvious properties that it should enjoy.
  One such property is monotonicity-that is, the result of the
  function should grow (or at least not decrease) as the argument is
  increased.  This property can be expressed as follows.

\begin{Verbatim}
  property("Monotonicity") =
   forAll ((x:Int, y:Int) =>
	x<=y ==> (sum(x) <= sum(y)))
\end{Verbatim}

  This property actually fails for our implementation.  In particular,
  we get the following output from ScalaCheck:

\begin{Verbatim}
  ! SmallExamples.Monotonicity: Falsified after 0 passed tests.                 
  > ARG_0: 0 (orig arg: -2147483648)
  > ARG_1: 65536 (orig arg: 660619302)
\end{Verbatim}

  ScalaCheck is saying that the property fails, and in particular that
  values 0 and 65536 for x and y demonstrate that this property fails.
  We can get a better idea of what this test has uncovered by
  inspecting the results that our sum function returns for these values:

\begin{Verbatim}
  println (sum(0))     // prints 0
  println (sum(65536)) // prints -2147450880
\end{Verbatim}

  By stating our monotonicity property, we have uncovered an overflow
  problem.  The test points us to the fact that the Int type is finite
  and uses a fixed number of bits to represent just a subset of the
  integers.  When we add one too many numbers, the result is a value
  that is not representable, and we get instead another, meaningless
  value.

  It is highly instructive to note that we had already expressed
  several properties up to this point, and they all passed the test
  and were all valid.  But they were not really enough to help us
  realize that our implementation had limitations that we may have not
  thought of before.  This example also illustrates that the
  properties we express may not always be saying precisely what we
  thought they were saying when we wrote them.  In this instance, part
  of the problem comes from the fact that we used the same language
  for both the program and the properties, and both including
  reference to the finite type Int and the same definitions for the
  arithmetic and comparison operators defined on that type.

\begin{exercise}
  A smart programming language that supports recursion would allow us
  to use a definition like (``Sum (B)'') directly as code, and so we can
  define our function simply by copying the above if statement into its
  body as follows:

\begin{Verbatim}
  def sum3 (n:Int) : Int = if (n<1) 0 else n + sum3 (n-1)
\end{Verbatim}

  This is a reasonable solution for the problem that we set out to
  solve.  However, while the reasoning with which we arrived at this
  solution is perfectly valid under the right assumptions, the fact
  that we have expressed the various intermediate properties that lead
  to this solution in an executable form allows us to use ScalaCheck
  to reveal to us that our assumptions were not completely valid.

  Use ScalaCheck to test that the above recursive definition for sum
  actually satisfies all of the properties that we described.  If any of
  the properties break, explain why, and show how to solve this
  problem.

\end{exercise}

\section{Functions on Aggregate Types (I)}

  The way we develop and think about programs is affected by the type
  of data structures on which they operate.  So, when we work with Int
  values, many of our properties will look a lot like the kinds of
  properties that we see in traditional math courses, such as
  symmetry, monotonicity, and so on.  When we work with other data
  structures, such as lists for example, there will still be
  interesting patterns to the mathematical properties that we use to
  describe the behavior of such programs, but these patterns may seem
  different from the ones that we have seen in high school math
  classes.  As we see more examples of using properties to describe
  and reason about programs, clear patterns will definitely emerge.

\subsection{Surface syntax for lists}

  Aggregates or collections are data structures that bring together
  several small units of data.  A basic data structure is the list.
  In Scala, we construct lists by writing something like List (1,2,3)
  to represent a list of three elements, all integers, and in this
  case consisting of the elements 1, 2, and 3.

\subsection{Deeper structure and inductive nature of lists}

  Aggregate types, including lists, are often inductively defined.
  Knowing that a type is inductively defined is very helpful both for
  writing programs that take values of this type as input and for
  thinking about the kinds of properties that such programs should
  have.  Write List (1,2,3) is in fact syntactic sugar for the more
  primitive list constructors, which would express the same list as
  1::(2::(3::Nil)), where Nil is the empty list, and the operator (::)
  is the constructor for non-empty lists, which takes an element and a
  list as an argument.  Knowing the names of these primitive
  constructors for lists allows us to use the Scala match statement to
  write a function that checks whether a list is empty or not:

\begin{Verbatim}
  def isEmpty (list : List[Int]) =
    list match { case Nil                => true
                 case number :: listRest => false }
\end{Verbatim}

  The match statement not only allows us to check which constructor
  was last used to build the list but also allows us to refer to the
  components of the non-empty constructor (by the names ``number'' and
  ``listRest'' in the second branch of the case if we want to).

\subsection{Use cases for a function on lists}

  With these primitives in hand, we can approach our first programming
  task relating to lists: develop a function ``count'' that takes an
  integer and a list of integers and returns a count of the number of
  times this integer occurs in the list.

  To check our understanding of the required functionality, we write
  down several examples.  In practice, writing down examples is a
  useful thing to do while still with the customer discussing
  requirements, and also at the start of programming to make sure that
  we have a clear understanding of the task at hand.

\begin{Verbatim}
  property ("Use cases for count") =
    (count (7, List()) == 0) &&
    (count (7, List (7)) == 1) &&
    (count (7, List (1,7)) == 1) &&
    (count (7, List (7,1,7)) == 2 )
\end{Verbatim}

\subsection{Defining a function on lists}

  To think clearly about a new function that we have been asked to
  write, it is actually very useful to keep in mind the properties of
  the type of value it takes as input.  If the input type is
  inductively defined, which is the case for lists, this generally
  means that we should use a match statement for case analysis and
  recursion that (often directly) mimics the pattern of recursion in
  the inductive definition of the type itself.  In the case of this
  task, we can implement the function count as follows:

\begin{Verbatim}
  def count (number : Int, list : List [Int]) : Int =
    list match
      {case Nil => 0
       case number2 :: listRest
       => if (number == number2) 1 + count (number, listRest)
                            else count (number, listRest)}
\end{Verbatim}

  The code above passes our test examples, so we can turn to thinking
  about some deeper properties that can give us further assurance
  about our implementation.

\subsection{A non-defining property}

  One property of this function is that count for a number in two
  appended lists should be the same as the sum of the count in each of
  the two lists individually.  The notation for appending lists is
  simply (++), so we can express this property as follows:

\begin{Verbatim}
  property ("Count/append") =
    forAll((number: Int,
            list1 : List[Int],
            list2 : List[Int]) => 
              count(number,list1) + count(number,list2) 
              == count(number, list1++list2))
\end{Verbatim}

  While the above property is highly generic, it can be satisfied by
  an implementation that is certainly not correct.  In particular, if
  we used a faulty implementation that always returns zero, it would
  also satisfy the property above.  To avoid this kind of problem, it
  is useful to add a property that captures our intuition that adding
  another instance of the number we are looking for into the list
  increases the count by one.  The following is an example of a
  property that captures this fact:

\begin{Verbatim}
  property ("Instance at start") =
    forAll((number: Int,
            list  : List[Int]) => 
              count(number,list) + 1
              == count(number, number :: list))
\end{Verbatim}

  Adding this property to our specification of the functionality of
  the counting function gives us much more information about our
  implementation.  But it is useful to realize that it is not a
  {\em complete} specification of counting.

\begin{exercise}
  Check that the following function satisfies both
  the ``Count/append'' and the ``Instance at start'' properties:

\begin{Verbatim}
  def mad_count (number : Int, list : List [Int]) : Int =
    list match
      {case Nil => 0
       case number2 :: listRest => 
        1 + mad_count (number, listRest)}
\end{Verbatim}

  If the function above satisfies those two properties, suggest
  at least one new function that also satisfies these two properties.

\end{exercise}

  One thing that the ``Instance at start'' property does not capture is
  that if the instance added to the start of the list is {\em not} the
  same number, then the total count should not change.

\begin{Verbatim}
  property ("Non-instance at start (A)") =
    forAll((number1: Int,
            number2: Int,
            list  : List[Int]) =>
              number1 !=number2
              ==> (count(number1,list)
                   == count(number1, number2 :: list)))
\end{Verbatim}

  If we rewrite this property to replace the implication with an if
  statement, it will be easier to see that this and the ``Instance at
  start'' property can in fact be combined:

\begin{Verbatim}
  property ("Non-instance at start (B)") =
    forAll((number1: Int,
            number2: Int,
            list  : List[Int]) =>
              if (number1 !=number2)
                  (count(number1,list)
                   == count(number1, number2 :: list))
              else true)
\end{Verbatim}

 And now we can replace the ``true'' branch with the property that we
 expect to hold when the two numbers are equal, and we have:

\begin{Verbatim}
  property ("Something at start (A)") =
    forAll((number1: Int,
            number2: Int,
            list  : List[Int]) =>
              if (number1 !=number2)
                  (count(number1,list)
                   == count(number1, number2 :: list))
              else (count(number1,list) + 1
                   == count(number1, number2 :: list)))
\end{Verbatim}

 We note well that we have ``+~1'' in the statement in the second branch
 of the if statement.

 Clearly, this property combines the power of our two past properties
 about what happens when we add an element to the list.  However, it
 still does not really fully specify the behavior of the counting
 functionality.  It may be surprising to the reader, but it still
 leaves an unbounded amount of freedom in the behavior of the function
 ``count''.  In particular, the following function would actually
 satisfy all of the properties that we have expressed so far:

\begin{Verbatim}
  def dracula (number : Int, list : List [Int]) : Int =
    list match
      {case Nil => 1
       case number2 :: listRest
       => if (number == number2) 
               1 + dracula (number, listRest)
          else dracula (number, listRest)}
\end{Verbatim}

 And, in fact, any version of the above function with any integer
 value in the case of the empty list (Nil) would have satisfied all
 of the properties that we have expressed so far.  We can incorporate this
 final, additional requirement into the last property by introducing
 an additional check for the case in which (list = Nil) and rewriting it
 as follows:

\begin{Verbatim}
  property ("Count specification (A)") =
    forAll((number1: Int,
            number2: Int,
            list  : List[Int]) =>
              (if (list == Nil) 
                    count (number1, list) == 0 
               else true)
              &&
              (if (number1 !=number2)
                   (count(number1,list) == 
                    count(number1, number2 :: list))
               else (count(number1,list) + 1 == 
                     count(number1, number2 :: list))))
\end{Verbatim}

 An interesting feature of this property is that it is a complete
 specification of our counting function.  In fact, it is surprisingly
 close in what it says to what the actual code of our implementation
 says in its text.

\begin{exercise}
  Explain the justification for equivalence of the following
  sequence of the properties to the last property stated above (``Count
  specification (A)''):

\begin{Verbatim}
  property ("Count specification (B)") =
    forAll((number1: Int,
            list  : List[Int]) =>
	      list match
                {case Nil => count (number1, list) == 0
                 case number2::listRest =>
                  if (number1 !=number2)
                       (count(number1,listRest)
                        == count(number1, list))
                  else (count(number1,listRest) + 1
                        == count(number1,  list))})

  property ("Count specification (C)") =
    forAll((number1: Int,
            list  : List[Int]) =>
	      list match
                {case Nil => count (number1, list) == 0
                 case number2::listRest =>
                  if (number1 == number2)
                       (count(number1,listRest) + 1
                        == count(number1, list))
                  else (count(number1,listRest)
                        == count(number1, list))})

  property ("Count specification (D)") =
    forAll((number1: Int,
            list  : List[Int]) =>
	      count (number1, list) == 
               (list match
                 {case Nil => 0
                  case number2::listRest =>
                       if (number1 == number2)
                            count(number1,listRest) + 1
                       else count(number1,listRest)}))
\end{Verbatim}
\end{exercise}

 It is both a good thing and a bad thing when the most intuitive
 properties that we can think of for our functions are essentially the
 same as the program that we use to implement it.  It's a good thing
 because it increases the chances that our implementation has these
 properties.  It can be a bad thing for a couple of reasons.  First,
 it could mean that we have not identified enough other
 properties of our function to allow us to double-check the core set
 of properties that we are using for defining this property.  Second,
 it can mean that we may have made the same mistake twice in
 both specifying and implementing our function.  The solution to the
 first problem is to continue thinking about properties that we would
 like our function to have.  An example of such a property is the
 ``Count/append'' property that came across at the start of this
 example.  The solution to the second problem can be to arrange for
 another team member who has not seen the code or solution to
 independently come up with the specification of the properties that
 they would like this function to have.

\section{Functions with an inverse}

  While we have barely scratched the surface of specifying properties of
  single functions, it is important to move on and to consider what
  happens when we are developing multiple functions that have
  interrelated properties.  There are numerous patterns of such
  interactions.  In this section, we will consider a simple example of
  a very common pattern, namely that of functions that have an
  inverse.  One of the nice features of this pattern is that it is
  easy to intuitively identify in many areas of computer science.
  Intuitively, anything that we can view as a kind of ``encoding'' is an
  example of such a property because we generally do not use that word
  unless there is both an encoding and a decoding function, and where
  the second function serves as an inverse to the first.

\subsection{Encoding integers as bits}

  As a simple example of this kind of pattern, we will consider the
  problem of converting from non-negative integers (natural numbers)
  into binary and back.  This problem can be informally specified as
  follows: if we are converting {\em decimal} numerals then we would be
  turning a value like 125 into List(5,2,1).  We actually reverse the
  order and have the least significant digit come first in the list
  because that makes the functions just a bit more convenient to
  write.  In any case, what we want here is a {\em binary} encoding.  So,
  for an integer like 6 that is represented in binary by 110 we will
  want to get back the List(0,1,1).  In fact, because with binary
  digits we only have two choices of digit, we want the result to use
  bools and represent 0 with false and 1 with true, so that the
  result for the example above is List(false, true, true).

\subsection{Use cases for encoding}

  As usual, we start with writing down examples, and we can write
  the function immediately after writing the examples.  In both of the
  following cases, ScalaCheck tells us that both functions have passed
  our tests with flying colors:

\begin{Verbatim}
  property("Use cases for encode") =
    (encode (0) == List ()) &&
    (encode (1) == List (true)) && 
    (encode (2) == List (false, true)) &&
    (encode (3) == List (true,  true)) &&
    (encode (8) == List (false, false, false, true))

  def encode (n : BigInt) : List[Boolean] =
   if (n <= 0)
    List ()
   else ((n % 2) == 1) :: (encode (n / 2))

  property("Use cases for decode") =
    (decode (List ()) == 0) &&
    (decode (List (true)) == 1) && 
    (decode (List (false, true)) == 2) &&
    (decode (List (true,  true)) == 3) &&
    (decode (List (false, false, false, true)) == 8)

  def decode (list : List[Boolean]) : BigInt =
   list match
     {case Nil => 0
      case b::listRest =>
          (decode (listRest)) * 2 + (if (b) 1 else 0)}
\end{Verbatim}

  In the case of the two functions above, the encoding and decoding
  functions seem so simple that it is hard to imagine that anything can
  be wrong with them.  However, even if the functions appear to
  provide precisely the functionality that we need, it is useful to
  explore the properties of these functions to make sure that we also
  understand the functionality that we have just provided.  As a first
  example, we will state the property that encoding followed by
  decoding produces the same value we started with.  That is,

\begin{Verbatim}
  property ("n>=0 ==> d(e(n)) == n") =
    forAll ((n : BigInt) => 
       n>=0 ==> (decode (encode (n)) == n))
\end{Verbatim}

  While this property holds for this function, it is useful to note
  that the dual property does not hold.  That is, decoding following by
  encoding does not produce the same binary digit.  This may seem a
  bit curious, but in fact this is a common situation for many
  encoding/decoding pairs.  

\subsection{The asymmetry of invertible functions}

  Expressing this property in ScalaCheck allows us to find a counter
  example:

\begin{Verbatim}
  property ("e(d(l)) == l (false!)") =
    forAll ((l : List[Boolean]) => 
       encode (decode(l)) == l)
\end{Verbatim}

  Testing this property quickly returns a counter example that gives
  us a good idea of why it does not hold:

\begin{Verbatim}
  ! SmallExamples.e(d(l)) == l (false!): Falsified after 6 passed tests.        
  > ARG_0: List("false")
\end{Verbatim}

  What has happened is that in our encoding the number 0 is
  represented by the empty list.  Decoding works fine on the empty
  list to produce 0.  Moreover, it will also produce that same value
  for any list that is all zeros.  But for all such lists, the encoding
  function will always be only the empty list, which is not really the
  same list as all the lists of all zeros.  In fact, in general,
  applying our decoding and encoding functions will always have the
  effect of removing all leading zeros.  And so they don't always
  give us back exactly what we started with, which is what the property
  above says.

  It is useful to note that decoding followed by encoding does give us
  back what we started with {\em if} what we started with was produced
  by the encoding function.  This can be expressed as follows:

\begin{Verbatim}
  property ("n>=0 ==> e(d(e(n))) == e(n)") =
    forAll ((n : BigInt) => 
       n>=0 ==> (encode (decode (encode (n))) == encode (n)))
\end{Verbatim}

  Thus, the key property of such encoding systems is that the encoding
  function is an inverse of the decoding function, which is what is
  captured by the first property.  This is a pattern that we will see
  in other computing applications that go by names such as encryption,
  serialization, parsing, and others.

\begin{exercise}
  Explain why it may not be particularly surprising if
  the property:

\begin{Verbatim}
    "n>=0 ==> d(e(n)) == n"
\end{Verbatim}

  then the property

\begin{Verbatim}
    "n>=0 ==> e(d(e(n))) == e(n)"
\end{Verbatim}

  also holds.
\end{exercise}

\begin{exercise}[Lab]
While it is the case that an encoding function is not always an
inverse to the decoding function, it is worthwhile to note that this
can be a useful property when it holds.  In particular, it means that
every target representation is a unique encoding of a number in our
source for the encoding.

Write the code and properties for an encoding/decoding pair from
integers to lists of booleans and where this symmetry does hold.
Hint: Treat the empty list as always holding an implicit true at the
most significant digit, and subtract one from the interpretation so
that 0 is still representable.
\end{exercise}

\begin{exercise}[Lab]
Define an addition function on the original encoding of integers.
State its correctness property, and check it using ScalaCheck.
\end{exercise}

\section*{Acknowledgments}  

We would like to thank the participants of the summer school that was
organized at Halmstad University from May 30th to June 1st.  A special
thanks also goes to the fourth speaker the summer school, Prof. John
Hughes.  The goals of the workshop would not have been achieved
without the active efforts of the faculty of Halmstad University that
worked closely with the authors to incorporate these ideas into the
undergraduate and graduate curricula, especially Nicolina Månsson,
Elisabeth Uhlemann, Ulf Holmberg, Tony Larsson, and Thorsteinn (Denni)
Rögnvaldsson.  Prof. Page's visit to Halmstad was greatly facilitated
by the hard work of Eva Nestius, Magnus Jonnson, Bertil Svensson, and
Magnus Larsson.  Finally, Paul Brauner kindly provided us with helpful
comments on a draft of these notes.

Rex's visit would have not been possible without the generous support
of the U.S. Department of State through the Fullbright Scholar
program, as well as the support of Halmstad University.

\addcontentsline{toc}{section}{Bibliographic References}
\bibliographystyle{eptcs} 
\bibliography{notes}

\begin{thebibliography}{1}
\providecommand{\bibitemdeclare}[2]{}
\providecommand{\urlprefix}{Available at }
\providecommand{\url}[1]{\texttt{#1}}
\providecommand{\href}[2]{\texttt{#2}}
\providecommand{\urlalt}[2]{\href{#1}{#2}}
\providecommand{\doi}[1]{doi:\urlalt{http://dx.doi.org/#1}{#1}}
\providecommand{\bibinfo}[2]{#2}

\bibitemdeclare{misc}{Accuracy}
\bibitem{Accuracy}
\emph{\bibinfo{title}{Accuracy and Precision}}.
\newblock \bibinfo{howpublished}{Wikipedia, The Free Encyclopedia}.
\newblock \bibinfo{note}{Available online from {\tt wikipedia.org}. Viewed May
  2011.}

\bibitemdeclare{book}{ByExample}
\bibitem{ByExample}
\bibinfo{author}{Gojko Adzic} (\bibinfo{year}{2011}):
  \emph{\bibinfo{title}{Specification by Example: How Successful Teams Deliver
  the Right Software}}.
\newblock \bibinfo{publisher}{Manning}, \bibinfo{address}{Greenwich, CT}.

\bibitemdeclare{inproceedings}{Test}
\bibitem{Test}
\bibinfo{author}{Robert Cartwright} (\bibinfo{year}{1981}):
  \emph{\bibinfo{title}{Formal Program Testing}}.
\newblock In: {\sl \bibinfo{booktitle}{Principles of Programming Languages}},
  pp. \bibinfo{pages}{125--132}, \doi{10.1145/567532.567546}.

\bibitemdeclare{inproceedings}{QuickCheck}
\bibitem{QuickCheck}
\bibinfo{author}{Koen Claessen} \& \bibinfo{author}{John Hughes}
  (\bibinfo{year}{2000}): \emph{\bibinfo{title}{{Q}uick{C}heck: A Lightweight
  Tool for Random Testing of {H}askell Programs}}.
\newblock In: {\sl \bibinfo{booktitle}{International Conference on Functional
  Programming}}, pp. \bibinfo{pages}{268--279}, \doi{10.1145/351240.351266}.

\bibitemdeclare{misc}{ScalaCheck}
\bibitem{ScalaCheck}
\emph{\bibinfo{title}{{ScalaCheck} Tutorial}}.
\newblock \bibinfo{howpublished}{Available online from {\tt
  http://code.google.com/p/scalacheck/wiki/UserGuide}}.
\newblock \bibinfo{note}{Viewed May 2011.}

\bibitemdeclare{misc}{UseCase}
\bibitem{UseCase}
\emph{\bibinfo{title}{{Use-case Analysis}}}.
\newblock \bibinfo{howpublished}{Wikipedia, The Free Encyclopedia.}
\newblock \bibinfo{note}{Available online from {\tt wikipedia.org}. Viewed May
  2011.}

\end{thebibliography}

\appendix
\section{Broader Educational Context}

The seed for this tutorial was planted when Rex organized a workshop
on Teaching Software Correctness in May 2008.  Walid attended the
workshop and was struck by the unusual effectiveness of property-based
testing in helping programmers develop a program and a property that can
actually be verified mathematically by a theorem prover.  While it was
impressive to see the theorem prover used in the workshop (ACL2)
to automatically prove a host of sophisticated properties, what was most
impressive was the effect that the combination of property-based
testing had {\em on the programmer}: property-based testing guided the
programmer to a correct program, which is really the only plausible
candidate for even starting to think about the completely independent,
non-trivial, and often quite labor-intensive task of {\em proving} a
program correct.

As luck would have it, two years later, in May 2010, Rex was
organizing the Symposium on Trends in Functional Programming, which
Walid attended.  At that meeting, Walid mentioned to Rex that he was
moving to Halmstad University, which is in the process of starting up
an independent Ph.D. program, and many of the faculty are genuinely
interested in curricular development.  Thanks to a Fulbright
Scholarship, exactly one year later, Halmstad University hosted Rex
for a one-month visit aimed at introducing his ideas into the Embedded
and Intelligent Systems curricula at both the graduate and
undergraduate levels.

Working on exploring how this can be achieved started by a kickoff
meeting that was organized the first week of the visit, which was well
attended by faculty.  It included a presentation of the bachelor's
program (three years) by Nicolina Månsson and of the masters program
(two years) by Jörgen Carlsson, followed by a presentation by Rex of
the key ideas in his approach.  Both in terms of research and
education, the strengths of Halmstad University's programs considered
to be:
\begin{itemize}
 \item Multicore architectures
 \item Real-time communications
 \item Cooperative embedded systems
 \item DSLs and program generation
 \item Modeling and simulation of cyberphysical systems
\end{itemize}
The effectiveness of both research and education of these areas would
be strengthened if students were more capable of producing
high-quality software.  It was therefore instructive to learn of the
results of the careful analysis that Rex had conducted of a wide range
of undergraduate curricula in the U.S., as well as his varied efforts
to introduce more ``property-based thinking'' into various courses.
For example, it seemed that students tend to find that introducing
testing-based methods into math classes makes them easier.  On the
other hand, introducing these ideas into software engineering gets
students' attention primarily because of its novelty.  Potential
employers of students seem to easily appreciate the value of getting
students who are trained to rigorously test their programs, and who
are capable of producing high-quality software.  It was also
interesting to learn from Rex that there are studies that show that
testing techniques based on analysis of the code itself are generally
much more economically viable than ones based solely on behavior.

Two general problems with computer science curricula were identified.
The first is that the theory classes, such as discrete math classes,
tend to be disconnected from the artifacts that students in this
discipline are most interested in, such as software and hardware.  The
other problem is that the courses that teach important theoretical
tools, such as logic for example, tend to focus on the meta-theory of
such tools rather than on how to use them in relation to concrete
artifacts that students are interested in (again, software and
hardware).  In what class do we learn how to use logical statements to
state mathematical properties of programs, not to mention learn to
reason about programs in terms of such properties?

It was suggested that the most practical and possibly most effective
approach will be to incorporate Rex's ideas into the curriculum by
injection into a variety of different courses.  Also, it was suggested
that using property-based testing could be an effective vehicle for
introducing a wide range of concepts relating to software correctness
and logic.  With such a foundation, students are prepared to read
books or follow courses that are more mathematically oriented, either
on their own or in the context of advanced courses offered at the
university.

Several candidate areas for introducing these ideas where identified.
In the master's degree program, where books and instruction are in
English:
  \begin{itemize}
	\item Embedded systems programming course (using ScalaCheck and maybe ACL2)
	\item Cooperating intelligent systems course
	\item Discrete math course
	\item Cyber-physical systems (CPS)
  \end{itemize}
Several specific ideas were discussed for CPS.  Natural candidates
included scheduling problems.  Some specific ideas: 1. number of steps
in a computation (e.g. interrupt handler), 2. model interrupt handler
as an ACL2 computation + step count, 3.  prove bound on the number of
steps, 4. example: moving window algorithms, 5. bounded work to
update, and 6. d.s. maintains an invariant.  All that would be needed
would be two examples of this, and to create material for a lecture +
homework.  This can be packaged as a one-week module, which may also
be usable in other courses.

In the bachelor's program, where books and notes are in Englsih but instruction
          is in Swedish:
   \begin{itemize}
	\item Programming I, teaching specifications
          together with programs.
	\item Programming II, teaching  contracts
	\item Switching theory and digital control.
	\item Algorithms and data structures
	\item Computer systems organization
   \end{itemize}
This suggested that there should be no shortage of opportunities for
introducing Rex's ideas at Halmstad.  With this, it was agreed that
Veronica, Rex, and Walid would focus on preparing materials for the
workshop that would serve as practical starting points for the faculty
at Halmstad to explore these possibilities more concretely.  These are
the notes presented in this document.

The rest of the story will depend on how well we succeed at
integrating these ideas into various courses in the Halmstad
curriculum!

\end{document}